\documentclass {PoS}
 \usepackage{epsf,epsfig}

\def\xb{\overline{x}}

\def\vk{{\bf k}_{\perp}}

\title{Pion pole and transversity effects in hard exclusive meson leptoproduction}

\ShortTitle{Pion pole and transversity effects in hard exclusive meson leptoproduction}

\author{\speaker{Sergey Goloskokov}\\
        Bogoliubov Laboratory of Theoretical
Physics, Joint Institute
for Nuclear Research,\\ Dubna 141980, Moscow region, Russia\\
        E-mail: \email{goloskkv@thsun1.jinr.ru}}

\abstract{We investigate exclusive electroproduction of vector and
pseudoscalar mesons at large photon virtuality $Q^2$. These
reactions were analyzed within the handbag approach where
amplitudes factorize into hard subprocesses and generalized parton
distributions (GPDs).

The essential role of transversity effects were found in
pseudoscalar and light vector meson leptoproduction. These
contributions  are important at low $Q^2$ and determined by
twist-3 effects accompanied by transversity GPDs. The transversity
contributions lead to large transverse cross sections for most
reactions of pseudoscalar meson leptoproduction. The transversity
effects in vector meson production are visible in spin
observables.

We consider spin effects in the $\omega$  and $\rho^0$
leptoproduction reactions. It is shown that the pion pole
contribution is very important in the $\omega$ production. Such
effects in the $\rho^0$ channel are much smaller.  Our results on
spin asymmetries and spin density matrix elements in these
reactions were found to be in good agreement with HERMES data.}

\FullConference{XXII International Baldin Seminar on High Energy Physics Problems,\\
        15-20 September 2014\\
        JINR, Dubna, Russia}

\begin{document}

\section{introduction}
Our investigations of hard electroproduction of vector mesons
\cite{gk06} were based on the handbag approach where the amplitude
of meson production at high $Q^2$ factorizes into hard meson
electroproduction off partons, and  GPDs  \cite{fact}.  The  hard
subprocesses were analysed within the modified perturbative
approach \cite{sterman} where we considered the quark transverse
degrees of freedom accompanied by the Sudakov suppressions. This
approach describes successfully the data on $\rho^0$ and $\phi$
electroproduction.  We discussed in section 2.

The amplitudes of the pseudoscalar meson (PM) leptoproduction  in
the leading twist approximation are sensitive to GPDs
$\widetilde{H}$ and $\widetilde{E}$.   It was found that these
contributions  were not sufficient to describe spin effects in the
PM production at sufficiently low $Q^2$ \cite{gk09}. To be
consistent with experiment,  essential contributions from the
transversity GPDs  $H_T$, $\bar E_T$ are needed \cite{gk11}.
Within the handbag approach the transversity GPDs go together with
the twist-3 meson wave function. We discuss in  section 3 the role
of transversity effects in the PM leptoproduction at HERMES and
CLAS energies. We show that the transversity GPDs lead to a large
transverse cross section for most reactions of the pseudoscalar
meson production \cite{gk11},  which exceed the leading twist
longitudinal cross section. The role of transversity effects were
analysed in the vector meson (VM) leptoproduction too.  The
transversity GPDs were found to be important in the spin density
matrix elements (SDMEs) and spin asymmetries of the VM
leptoproduction with a transversely polarized target. The obtained
results are in good agreement with CLAS, HERMES and COMPASS data.

The  HERMES data on  SDMEs for the $\omega$ production indicated
strong contributions from unnatural parity exchanges. Using GPDs
from our analyses of the hard meson leptoproduction we
investigated  $\omega$ SDMEs \cite{gk14}. It was found that the
pion pole (PP) contribution plays an important role in the
$\omega$ production and is essential in explanation of the large
unnatural-parity effects observed by HERMES. Based on our approach
we found $\omega$ SDMEs to be in good agreement with the HERMES
experimental results \cite{omega14}. The PP contribution to the
$\rho^0$ production is much smaller with respect to the $\omega$
case. We discuss the PP effects in section 4.

\section{Handbag approach. Vector meson leptoproduction}
Within the handbag approach the meson production amplitude  at
sufficiently high photon virtuality $Q^2$ is factorized
\cite{fact} into a hard subprocess amplitude ${\cal H}$ and  GPDs
$F$ which contain information on the hadron structure. In the
forward limit and zero skewness GPDs are equivalent to ordinary
Parton Distribution Functions (PDFs). With the help of sum rules
they are connected with hadron form factors, and information on
the parton angular momenta can be extracted.

The leading contributions to the meson production amplitude off
non-flip proton can be described in terms of various parton
effects
\begin{equation}\label{ff}
M_{\mu' +,\mu +} \propto \int_{-1}^1 dx
   {\cal H}^a_{\mu' +,\mu +} F^a(x,\xi,t).
\end{equation}
Here $a$ is a flavor factor, ${\cal H}^a_{\mu' +,\mu +}$ is a hard
meson electroproduction amplitude off partons with the same
helicities, $\mu$ and $\mu'$ are helicities of the photon and
produced meson. In the VM production we have $F^a$ GPDs
contributions from gluons, quarks and sea. In the PM production
polarized GPDs $\tilde F^a$ give contribution.

The subprocess amplitude is calculated within the modified
perturbative approach \cite{sterman}. We consider the
$k_\perp^2/Q^2$ corrections in the propagators of the hard
subprocess amplitude ${\cal H}$ together with the nonperturbative
$\vk$-dependent meson wave function \cite{koerner}. The power
corrections can be regarded as an effective consideration of the
higher twist effects. The gluonic corrections are treated in the
form of the Sudakov factors whose resummation can be done in the
impact parameter space \cite{sterman}.

The GPDs are estimated using the double distribution
representation \cite{mus99}

\begin{equation}\
F_a(\xb,\xi,t) = \int_{-1}^1\,d \beta\,
\int_{-1+|\beta|}^{1-|\beta|}\,d \alpha
\,\delta(\beta+\xi\,\alpha-\bar x)\,f_a(\beta,\alpha,t).
\end{equation}
which connects  GPDs $F$ with PDFs $h$ through the double
distribution function
\begin{equation}
f_a(\beta,\alpha,t)= h_a(\beta,t)\,
                   \frac{\Gamma(2n_i+2)}{2^{2n_i+1}\,\Gamma^2(n_i+1)}
                   \,\frac{[(1-|\beta|)^2-\alpha^2]^{n_i}}
                           {(1-|\beta|)^{2n_i+1}}.
\end{equation}
Here n=1 for valence quarks and n=2 for gluon and sea
contributions.

 The  $t$- dependence in PDFs $h$ is considered in the Regge form
\begin{equation}\label{pdfpar}
h(\beta,t)= N\,e^{b_0 t}\beta^{-\alpha(t)}\,(1-\beta)^{n},
\end{equation}
and $\alpha(t)$ is the corresponding Regge trajectory.  The
parameters in (\ref{pdfpar}) are obtained from the known
information about PDFs \cite{CTEQ6} e.g, or from the nucleon form
factor analysis \cite{pauli}.

\begin{figure}[h!]
\begin{center}
\begin{tabular}{cc}
\includegraphics[width=7.2cm,height=5.2cm]{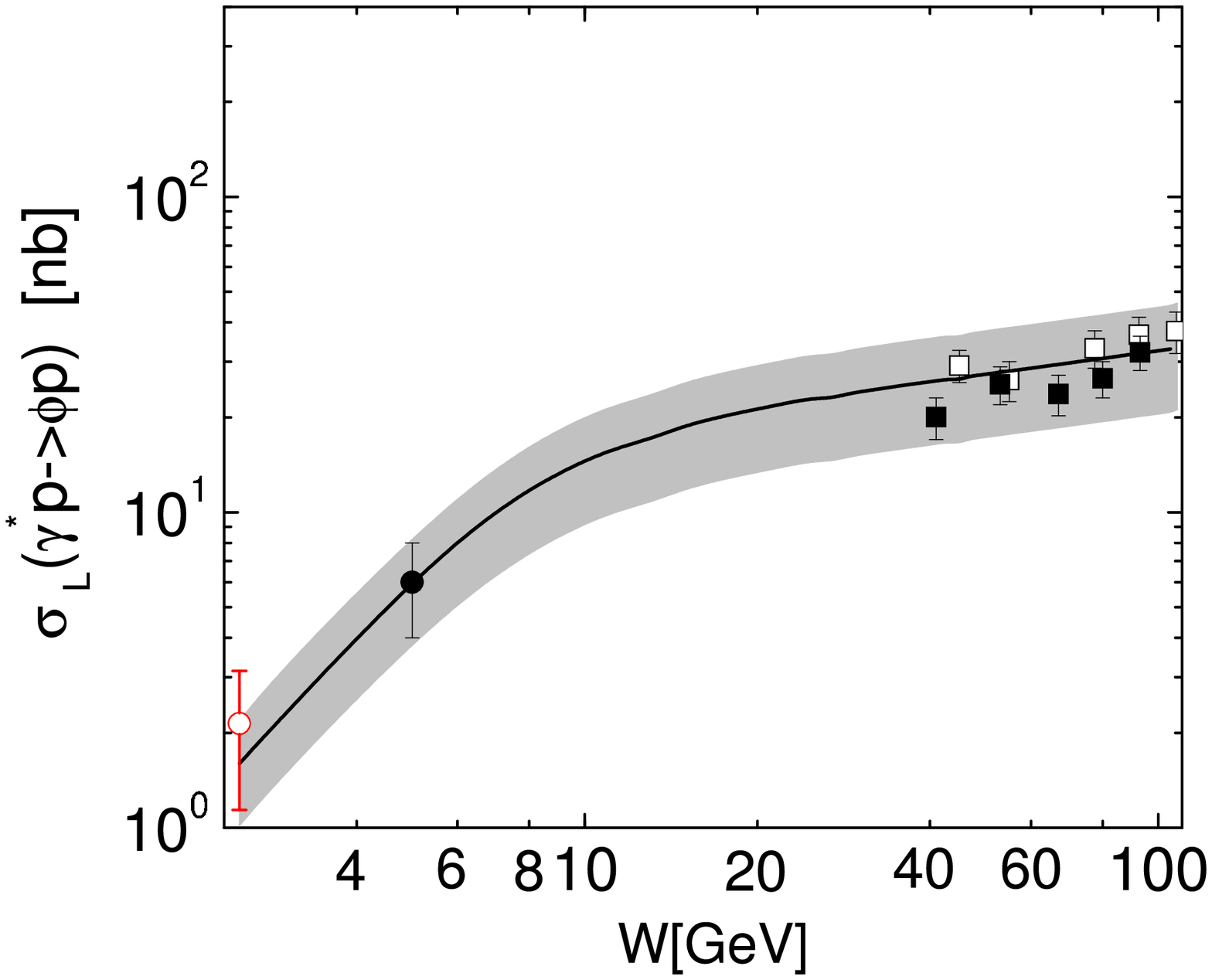}&
\includegraphics[width=7.2cm,height=5.2cm]{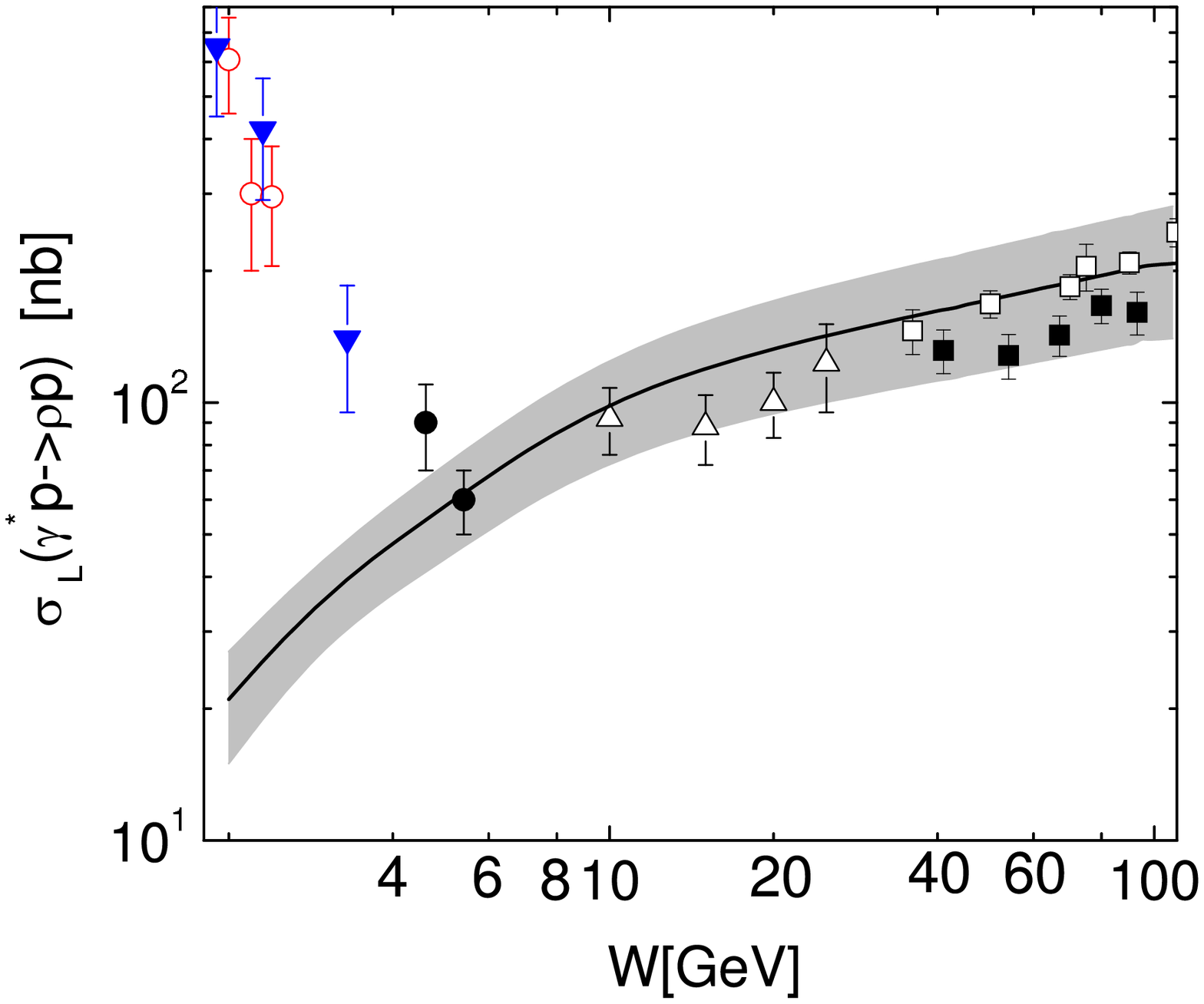}
\end{tabular}
\end{center}
\caption{Left: The longitudinal cross section  for $\phi$ at
$Q^2=3.8\,\mbox{GeV}^2$. Data: HERMES  (solid circle), ZEUS (open
square), H1 (solid square), open circle-  CLAS data point. Right:
The longitudinal cross section for $\rho$ production  at
$Q^2=4.0\,\mbox{GeV}^2$. Data: HERMES (solid circle), ZEUS (open
square), H1 (solid square), E665 (open triangle), CLAS- open
circles, CORNEL -solid triangle}
\end{figure}

The handbag approach was successfully applied to light meson
leptoproduction \cite{gk06}. The cross sections and spin
observables of light VM leptoproduction were found to be in good
agreement with HERMES, COMPASS and HERA data. As an example, in
Fig. 1, (left), we show our results for $W$ dependence of the
$\phi$ leptoproduction at fixed $Q^2$ which reproduce  the
experimental data in the whole range from CLAS to HERA energies.
This shows that gluon and sea GPDs work well from small to large
$x$- Bjorken. In Fig. 1, (right), we show a similar plot for the
$\rho$ production. We see that the model describes the $\rho$
meson leptoproduction quite well for $W> 4\mbox{GeV}$. The rapid
growth of the cross section at lower energies has not been
understood within the handbag model.

\section{Transversity effects in light meson leptoproduction.}
Exclusive electroproduction of PM was studied  within the handbag
approach \cite{gk09, gk11}. At the leading-twist accuracy the PM
production  is only sensitive to  GPDs $\widetilde{H}$ and
$\widetilde{E}$ which contribute to the amplitudes for
longitudinally polarized virtual photons \cite{gk09}. Such
contributions are not sufficient to describe the experimental
results on electroproduction of PM at low $Q^2$. We can show this
from the $A_{UT}^{\sin(\phi_s)}$ asymmetry, Fig. 2.

\begin{figure}[h!]
\begin{center}
\includegraphics[width=7.2cm,height=5.2cm]{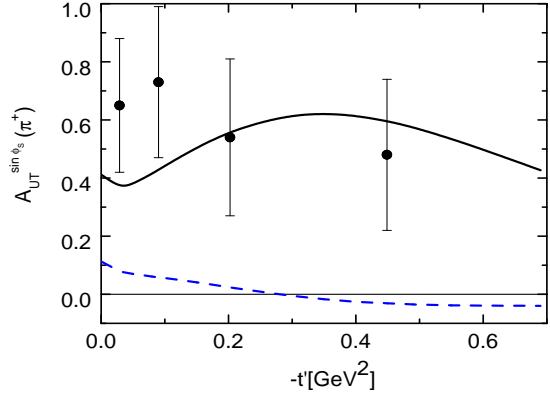}
\end{center}
\caption{ $A_{UT}^{\sin(\phi_s)}$ asymmetry: solid line prediction
of our approach. Dashed line- without twist-3 effects.}
\end{figure}
This asymmetry is proportional to $ M_{0-,++}$ and $M_{0+,0+}$
interference

\begin{equation}\label{autsfs}
A_{UT}^{\sin(\phi_s)} \propto \mbox{Im}[ M^*_{0-,++} M_{0+,0+}].
\end{equation}
In the handbag approach the amplitude $M_{0-,++} \propto  t'$.
Small PP contribution to $M_{0-,++}$ can not explain this
asymmetry- (dashed line in Fig. 2). Thus, the new sufficiently
large contribution to the $M_{0-,++}$ amplitude is needed.

It was found that at low $Q^2$ the PM leptoproduction data  also
require  contributions from  transversity GPDs  $H_T$ and $\bar
E_T$ which determine the amplitudes $M_{0-,++}$ and $M_{0+,++}$,
respectively. Within the handbag approach the transversity GPDs
are accompanied by a twist-3 meson wave function in the hard
amplitude ${\cal H}$ \cite{gk11} which is the same for both the
$M^{M,tw-3}_{0\pm,++}$ amplitudes
\begin{equation}\label{ht}
M^{M,tw-3}_{0-,++} \propto \,
                            \int_{-1}^1 d\xb
   {\cal H}_{0-,++}(\xb,...)\,H^{M}_T;\;
   M^{M,tw-3}_{0+,++} \propto \, \frac{\sqrt{-t'}}{4 m}\,
                            \int_{-1}^1 d\xb
 {\cal H}_{0-,++}(\xb,...)\; \bar E^{M}_T.
\end{equation}

The $H_T$ GPDs in the forward limit and $\xi=0$ are equal to
transversity PDFs $\delta$. We parameterize the PDF $\delta$ by
using the model \cite{ans}
\begin{equation}
  H^a_T(x,0,0)= \delta^a(x);\;\;\; \mbox{and}\;\;\;
\delta^a(x)=C\,N^a_T\, x^{1/2}\, (1-x)\,[q_a(x)+\Delta q_a(x)].
\end{equation}

The information on $\bar E_T$ is obtained now only in the lattice
QCD \cite{lat}. The lower moments of $\bar E_T^u$ and $\bar E_T^d$
were found to be quite large, have the same sign and a similar
size. We parameterize $\bar e_T$ PDF in the form (\ref{pdfpar})
with parameters determined from the lattice estimations.  The
double distribution  is used to calculate GPDs as before. Note
that  $H_T^u$  and $H_T^d$ GPDs have different signs. These
properties of GPDs provide an essential compensation of the $\bar
E_T$ contribution in the $\pi^+$ amplitude, but $H_T$ effects are
not small there. For the $\pi^0$ production we have the opposite
case -- the $\bar E_T$ contributions are large and the $H_T$
effects are smaller.

We present here our results on the PM  leptoproduction based on
the handbag approach. In calculation, we use the leading
contribution together with the transversity effects (\ref{ht})
which are essential at low $Q^2$.
 In Fig. 3 (left), we present the model results for the $\pi^0$
 production cross section \cite{gk11}.
  At small momentum transfer the $H_T$ contribution is
visible and provides a nonzero cross section. At $-t' \sim 0.2
\mbox{GeV}^2$ the $\bar E_T$ contribution becomes essential and
gives a maximum in the cross section. A similar contribution from
 $\bar E_T$ is observed in the interference cross section
$\sigma_{TT}$. The fact that we describe well both unseparated
$\sigma$ and $\sigma_{TT}$ cross sections can indicate that
transversity effects were probably observed in CLAS \cite{bedl}.
In Fig. 3 (right), we show  the $\eta$ and $\pi^0$ cross section
ratio obtained in the model  \cite{gk11}. At small momentum
transfer this ratio is controlled by the $H_T$ contribution. At
larger $-t$ the $E_T$ contributions become important. The value
about 1/3 for the cross section ratio in the momentum transfer
$-t'> 0.2\mbox{GeV}^2$ is a consequence of the flavor structure of
the $\eta$ and $\pi^0$ amplitudes. This result was confirmed by
the preliminary CLAS data \cite{vkubar}.
\begin{figure}[h!]
\begin{center}
\begin{tabular}{cc}
\includegraphics[width=7.2cm,height=5.2cm]{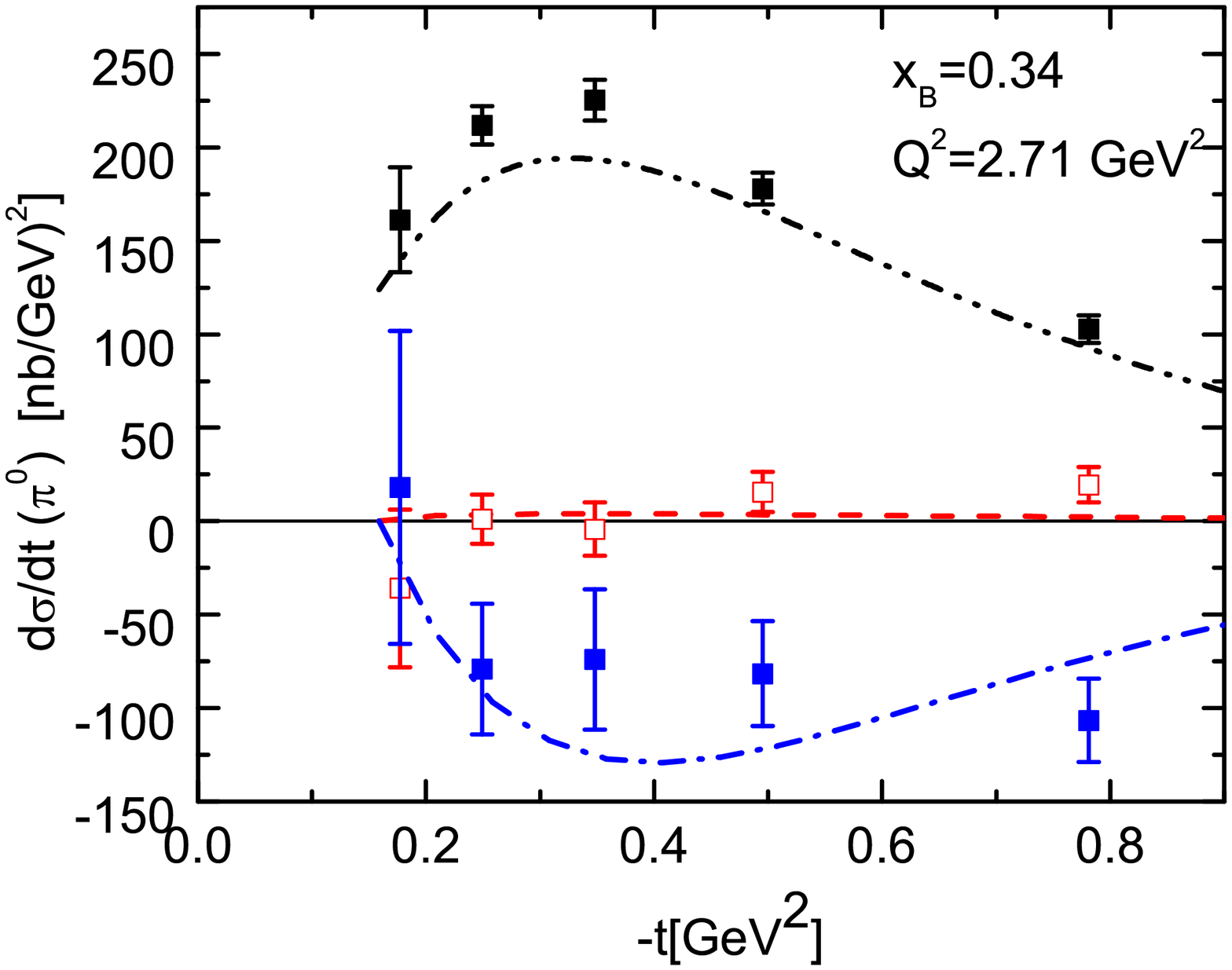}&
\includegraphics[width=7.2cm,height=5.2cm]{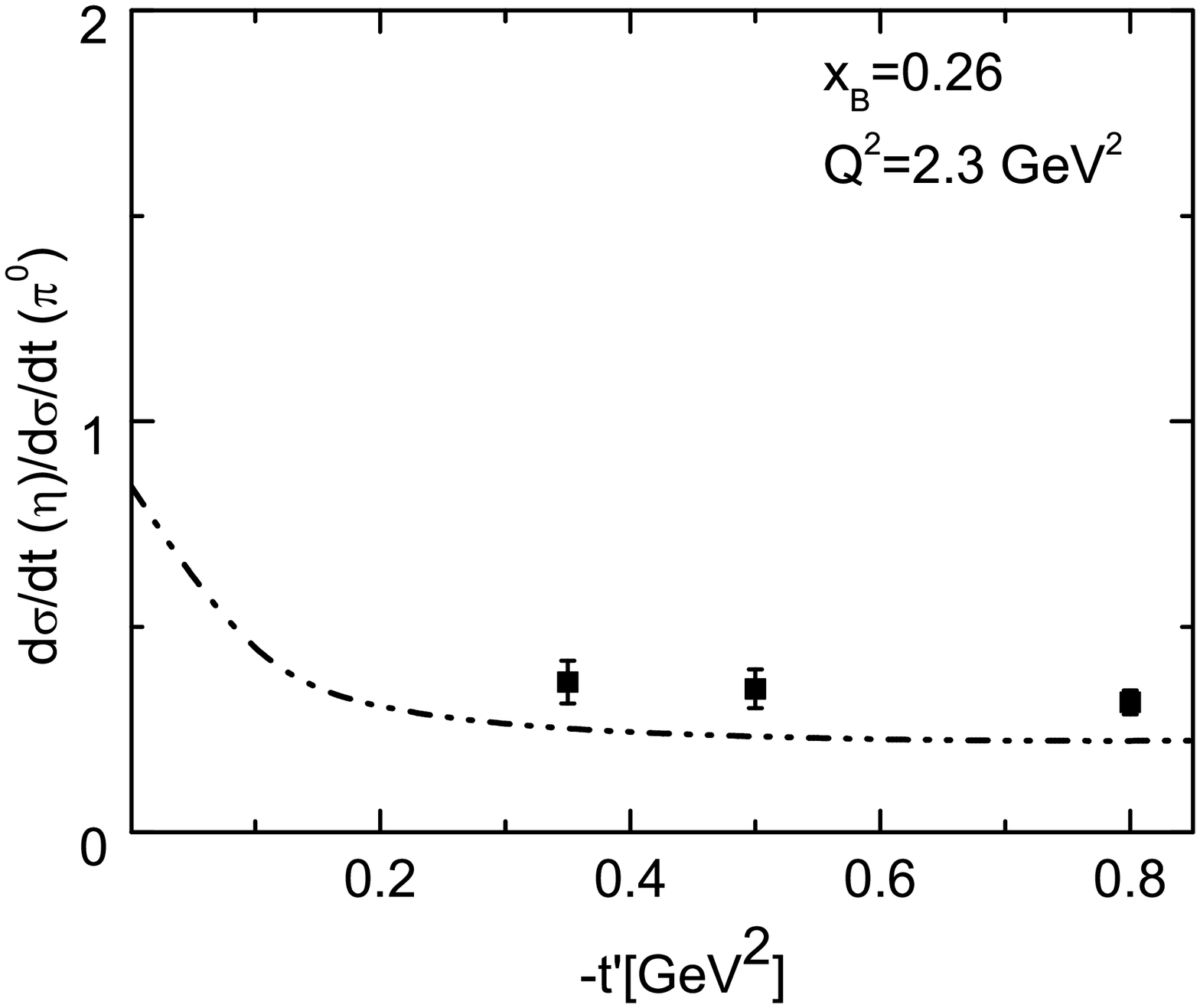}
\end{tabular}
\end{center}
\caption{Left: $\pi^0$ production in the CLAS energy range
together with the data \cite{bedl}. Dashed-dot-dotted line-
$\sigma=\sigma_{T}+\epsilon \sigma_{L}$, dashed
line-$\sigma_{LT}$, dashed-dotted- $\sigma_{TT}$. Right:
$\eta/\pi^0$ production ratio in the CLAS energy range together
with preliminary data \cite{vkubar}.}
\end{figure}

Now we  show some our results for transversity effects in the VM
leptoproduction. In this case, the transversity $M_{0-,++}$ and
$M_{0+,++}$ amplitudes have the form (\ref{ht}) but they are
parametrically about 3 times smaller \cite{gk13} with respect to
the PM amplitudes. In calculations,  we use the same
parameterizations for transversity GPDs $H_T$ and $\bar E_T$ which
were obtained in our study of the PM leptoproduction and can be
found in \cite{gk11,gk13}.

\begin{figure}[h!]
\begin{center}
\begin{tabular}{cc}
\includegraphics[width=7.2cm,height=5.8cm]{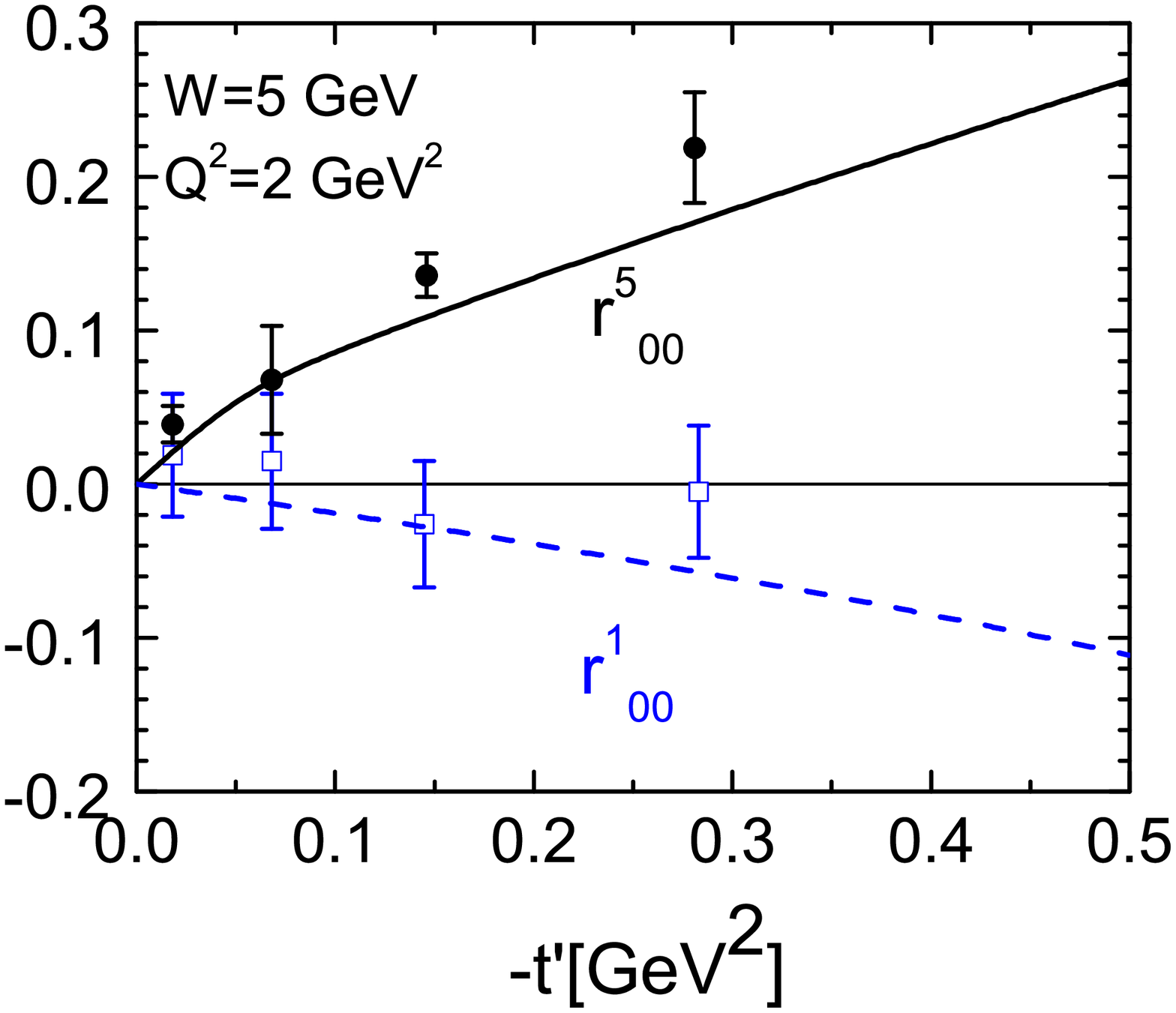}&
\includegraphics[width=7.2cm,height=5.8cm]{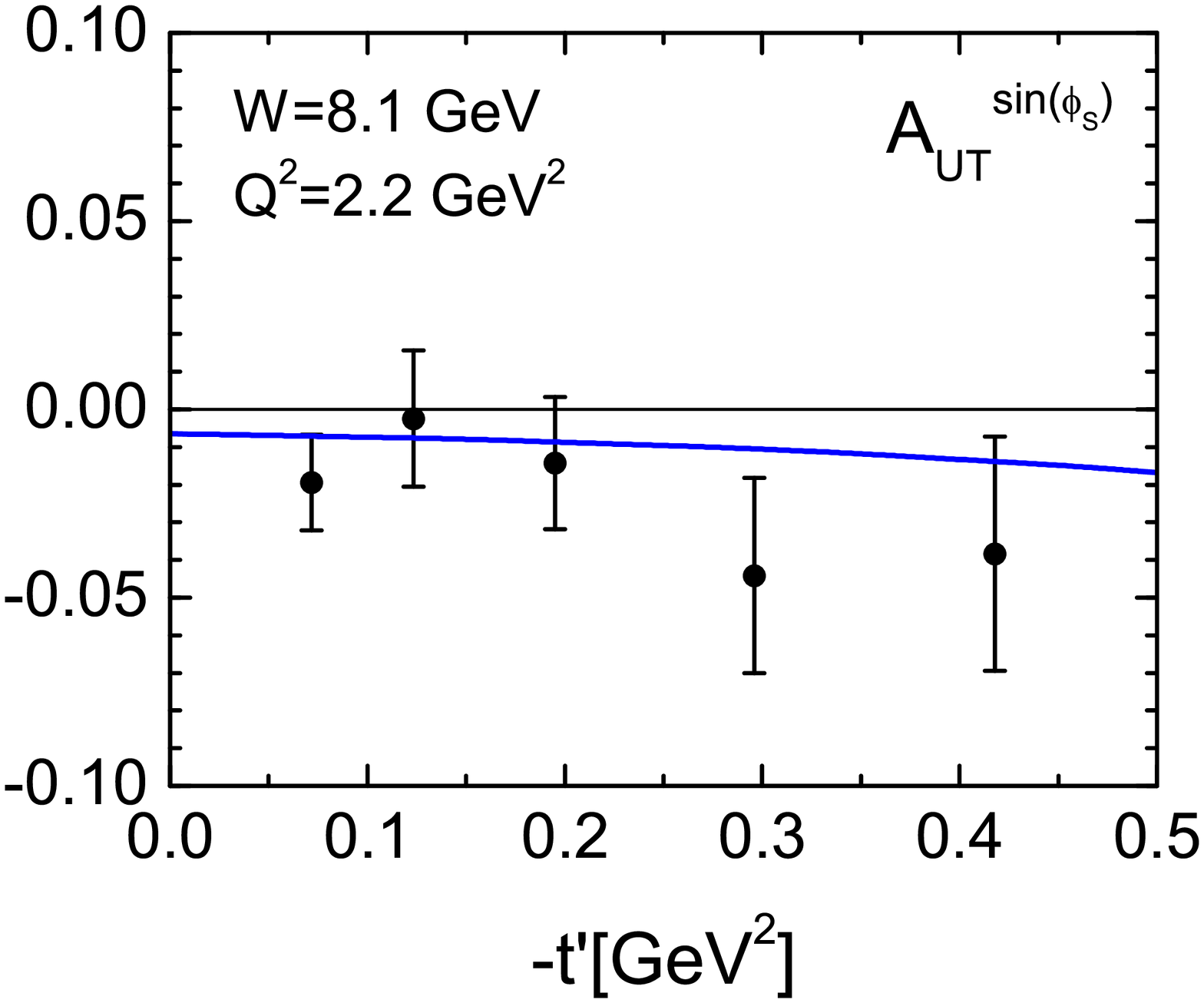}
\end{tabular}
\end{center}
\caption{Left:Transversity effects at SDMEs at $W=5 \mbox{GeV}$
together with HERMES data \cite{airap}. Right; Transversity
effects in $A_{UT}^{\sin(\phi_s)}$ moment of asymmetry at COMPASS
together with data \cite{autcomp13}}
\end{figure}

 The importance of the transversity GPDs was examined
in  the  SDMEs and  asymmetries measured with a transversely
polarized target. The  $\bar E_T$ contribution is essential in
some SDMEs. Really,
\begin{equation}\label{sdme}
r^5_{00} \sim \mbox{Re}[M_{0+,0+}^* M_{0+,++}];\;\;\; r^1_{00}
\sim -|M_{0+,++}|^2;\;\;\;  M_{0+,++}=<\bar E_T>.
\end{equation}
Our results \cite{gk13} for these SDMEs are shown in Fig.4,
(left). They reproduce HERMES  data \cite{airap} well.

 The $\sin (\phi_s)$ moment of the $A_{UT}$ asymmetry is determined by
the $H_T$ GPDs.

\begin{equation}\label{sinfs}
 A_{UT}^{\sin(\phi_s)} \sim \mbox{Im}[M_{0-,++}^*
M_{0+,0+}]; \;\;\;M_{0-,++}=<H_T>
\end{equation}

This asymmetry is found to be not small at COMPASS  \cite{gk13}
and compatible with the data \cite{autcomp13}, Fig. 4, (right).
Thus, we see that the transversity effects are important in the
description of PM and VM production at CLAS, HERMES and COMPASS
energy. Our results are compatible with experimental data.

\section{Large unnatural parity effects in $\omega$ production. Pion pole effects.}
The  HERMES data on the spin density matrix element for the
$\omega$ production indicate strong contributions from unnatural
parity contribution. The  natural and unnatural parity amplitudes
can be determined as
\begin{eqnarray}\label{mn}
M^N_{\mu' \nu',\mu \nu}=\frac{1}{2}\,\large [ M_{\mu' \nu',\mu
\nu}+(-1)^{\mu-\mu'}\, M_{-\mu' \nu',-\mu \nu} \large ],\nonumber\\
M^N_{\mu' \nu',\mu \nu}=\frac{1}{2}\,\large [ M_{\mu' \nu',\mu
\nu}-(-1)^{\mu-\mu'}\, M_{-\mu' \nu',-\mu \nu}\large ].
\end{eqnarray}

In most reactions the unnatural parity (UP) contributions are
small with respect to the natural one. However, in the $\omega$
production at HERMES \cite{omega14} it was found an unusual
result: the ratio of the unnatural to the natural parity cross
section
\begin{equation}\label{un}
U_1=2 \frac{d \sigma ^U(\gamma^*_T \to V_T)+\epsilon d \sigma
^U(\gamma^*_L \to V_L)}{d \sigma},
\end{equation}
which was expected to be small, was found to be larger than unity,
see Fig.5, (left).
\begin{figure}[h!]
\begin{center}
\begin{tabular}{cc}
\begin{minipage}{8.cm}
\includegraphics[width=7.cm,height=5.8cm]{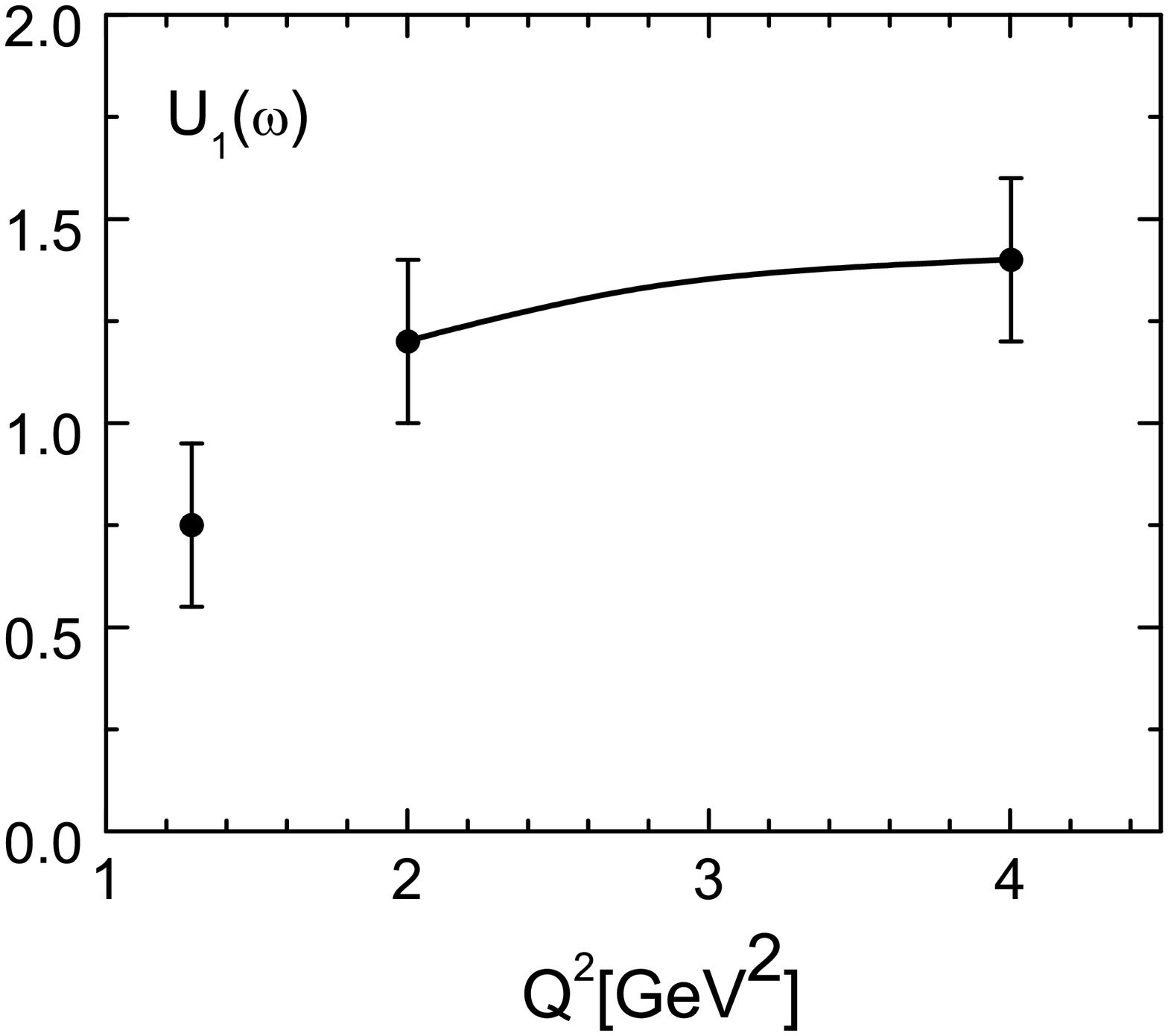}
\end{minipage}&
\begin{minipage}{8.cm}\phantom{.}\vspace{-10mm}
\mbox{\epsfig{figure=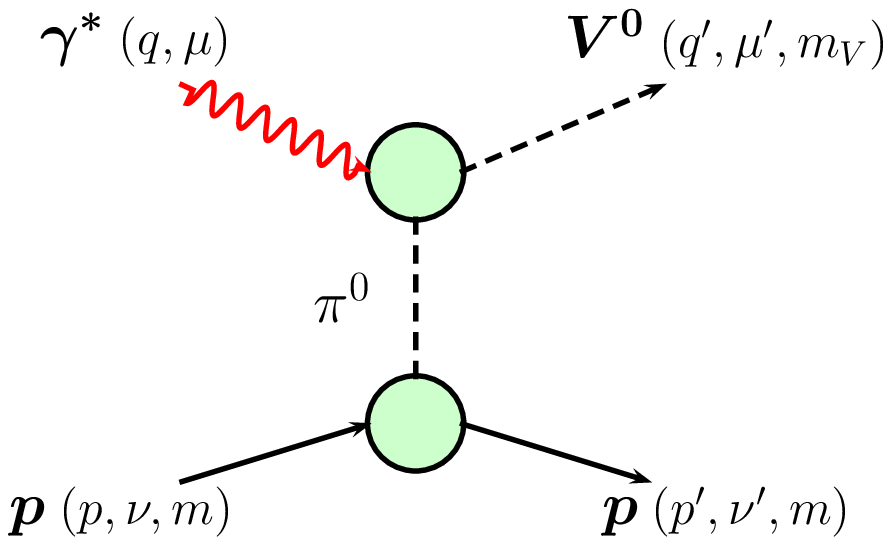,
width=6.2cm,height=3.5cm}}\end{minipage}
\end{tabular}
\end{center}
\caption{Left: $U_1$-the unnatural to natural parity cross section
ratio at HERMES \cite{omega14}. Right: the PP contribution in the
$\omega$ production.}
\end{figure}

Using  GPDs from our analyses of hard meson leptoproduction we
investigate \cite{gk14}   $\omega$ SDMEs  measured by the HERMES
Collaboration \cite{omega14}. It was found that the PP
contribution, Fig.5, (right), which has unnatural parity nature
gives an important effect in the $\omega$ production. The PP
contribution to helicity amplitudes is controlled by the $V
\pi^0\gamma$ transition form factor. The PP contribution to
helicity ampitudes looks like as follows:
\begin{equation}\label{omegap}
M^{pole}_{++,++} \sim \frac{\rho_{\pi V}}{t-m^2_\pi}\,
\frac{m\,\xi\, Q^2}{\sqrt{1-\xi^2}};\;\;\;\;\;\; M^{pole}_{+-,++}
\sim -\frac{\rho_{\pi V}}{t-m^2_\pi}\,\frac{\sqrt{-t'}\,Q^2}{2}
\end{equation}
with
\begin{equation}
\rho_{\pi V} \sim g_{\pi V}(Q^2)\, g_{\pi N\,N}\,F_{\pi N\,N}(t).
\end{equation}
The transition form factor $g_{\pi V}(0)$ is determined from the
VM radiative decay
\begin{equation}
\Gamma(V \to \pi \gamma) \sim \frac{\alpha_{elm}}{24}\, |g_{\pi
V}(0)|^2M_V^3.
\end{equation}
We find
\begin{equation}
|g_{\pi \omega}(0)|=2.3 \mbox{GeV}^{-1}; \;\;\;\;\;\;|g_{\pi
\rho}(0)|=.85\mbox{GeV}^{-1}.
\end{equation}
This means that $|g_{\pi \omega}|$ is about 3 times larger with
respect to $|g_{\pi \rho}|$ and we should observe large PP effects
in $\omega$ and small in $\rho$ production. The $Q^2$ dependence
of $g_{\pi V}(Q^2)$ was  extracted  from the $U_1$ data at $Q^2 <
4\mbox{GeV}^2$, Fig.5 (left).

In what follows we will discuss a comparison of our results on PP
effects in the $\omega$ and $\rho$ production with data at HERMES
energy. The natural and unnatural parity asymmetry $P$

\begin{equation}
P=\frac{d \sigma^N (\gamma^*_T \to V_T)-d \sigma^U (\gamma^*_T \to
V_T)}{d \sigma^N (\gamma^*_T \to V_T)+d \sigma^U (\gamma^*_T \to
V_T)}
\end{equation}
is an important example. If the UP contribution is small, we find
$P \sim 1$. If it is large, we have rather a different value for
the $P$ asymmetry.  We find that  with the PP contribution the
asymmetry $P \sim -0.5$ (full line) in agreement with experiment.
While neglecting the PP contribution we  obtain $P \sim 0.5$
(dashed line). Our results together with the HERMES data for
$\omega$ are shown in Fig.6, (left). In this figure we show for
comparison the model results for CLAS energy $W=3.5 \mbox{GeV}$ by
the dotted line and for COMPASS energy $W=8 \mbox{GeV}$ by the
dash-dotted curve. It can be seen that at COMPASS energies PP
effects are rather small for $\omega$. For the $\rho$ production
PP effects are small and the asymmetry $P$ is close to unity -see
Fig.6, (right).

\begin{figure}[h!]
\begin{center}
\begin{tabular}{cc}
\includegraphics[width=7.2cm,height=5.8cm]{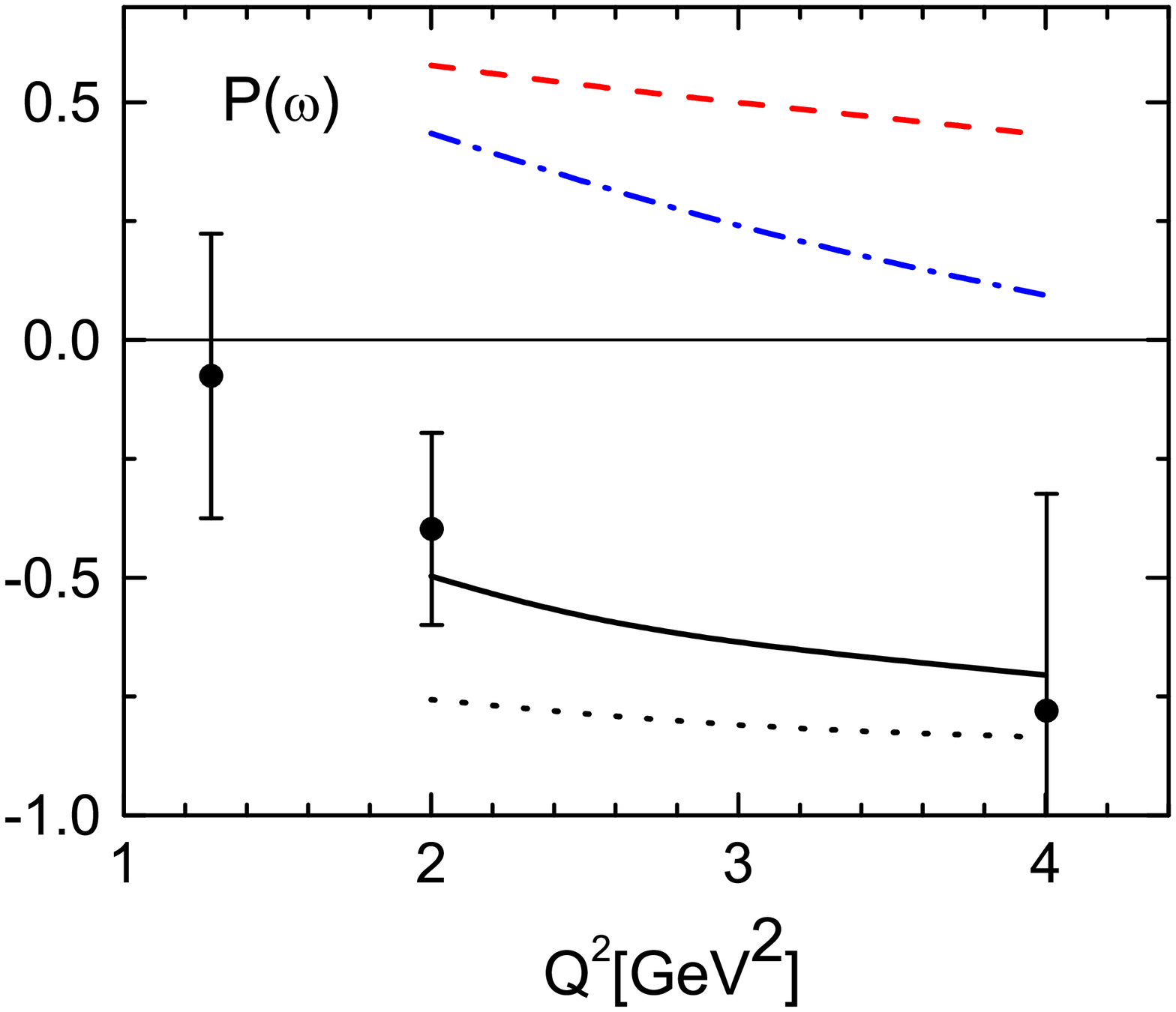}&
\includegraphics[width=7.2cm,height=5.8cm]{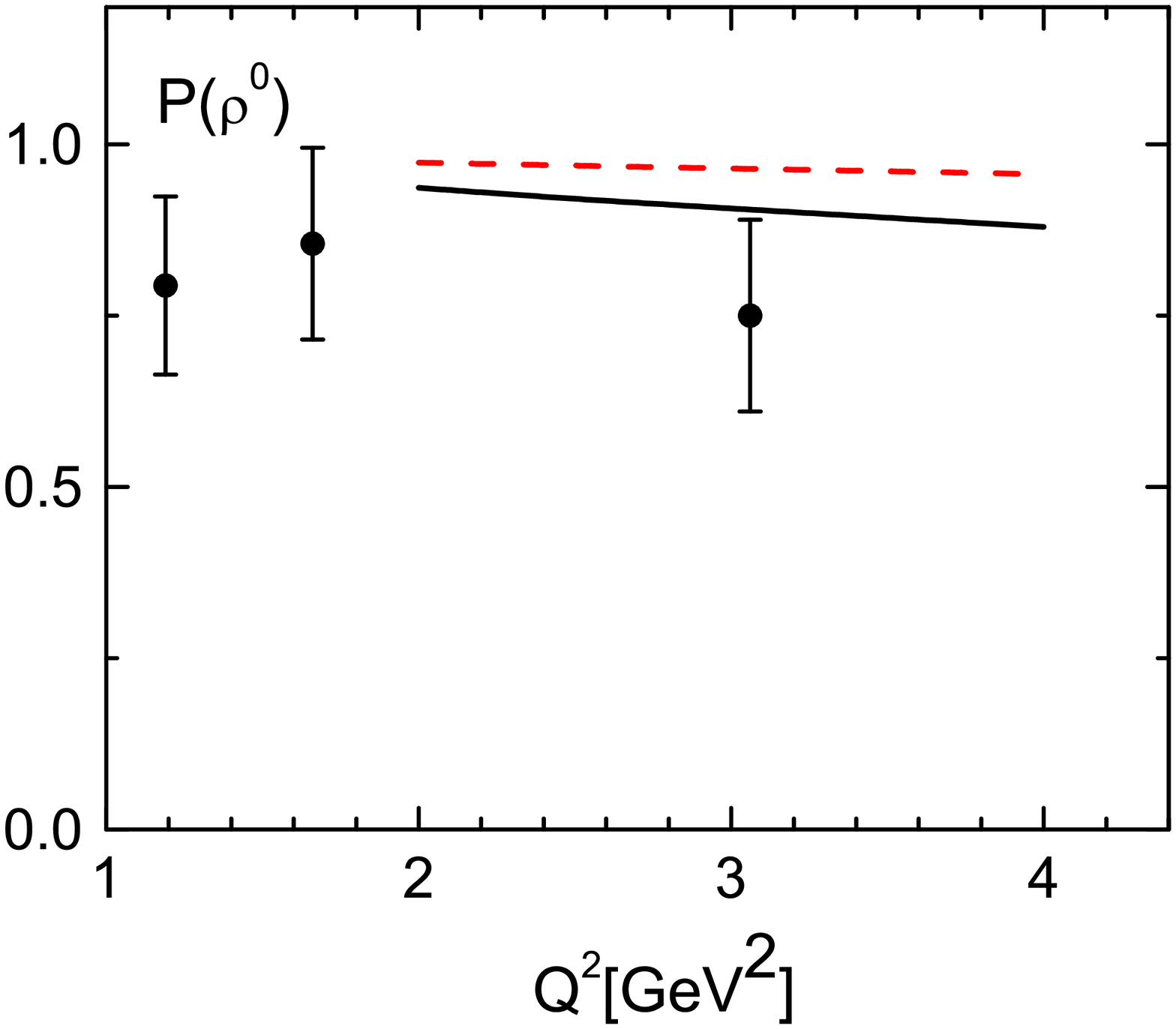}
\end{tabular}
\end{center}
\caption{Left: $P(\omega)$ asymmetry at HERMES. Black  solid line-
with PP, Red-dashed line -without PP. Black dotted line-for $W=3.5
\mbox{GeV}$ (CLAS), Blue dashed-dotted line for $W=8 \mbox{GeV}$
(COMPASS). Right: $P(\rho^0)$ at HERMES. Black solid with PP,
Red-dashed -without PP.}
\end{figure}

Interesting effects are observed in the ratio of the longitudinal
and the transverse cross section
\begin{equation}
R\simeq\frac{d \sigma(\gamma^*_L \to V_L)}{d \sigma(\gamma^*_T \to
V_T)}\nonumber
\end{equation}
 for the $\omega$ and $\rho$ production. The PP give an
 essential contribution to amplitudes with transversely polarized
 protons (\ref{omegap}). As a result, the PP effects lead
 to a quite small $R \sim 0.3$ ratio for the $\omega$ production at
 HERMES. This ratio is about $R \sim 1.2$ for the
 $\rho$ production  and close to the $\omega$ case without the PP
 contribution, see Fig. 7, (left). In
 this figure, we show for comparison our results for the  $R$ ratio for the $\omega$
production at CLAS and COMPASS energies.

The model  results for SDME
\begin{equation}
 r^{04}_{00} = \frac{d \sigma(\gamma^*_T \to V_L)+\epsilon
d \sigma(\gamma^*_L \to V_L)}{d \sigma}
\end{equation}
 for the $\omega$ production are shown in Fig. 7, (right). The
$r^{04}_{00}$ SDME is connected with the $R$ ratio and we find
similar results for both observables, see Fig 7.

\begin{figure}[h!]
\begin{center}
\begin{tabular}{cc}
\includegraphics[width=7.2cm,height=5.8cm]{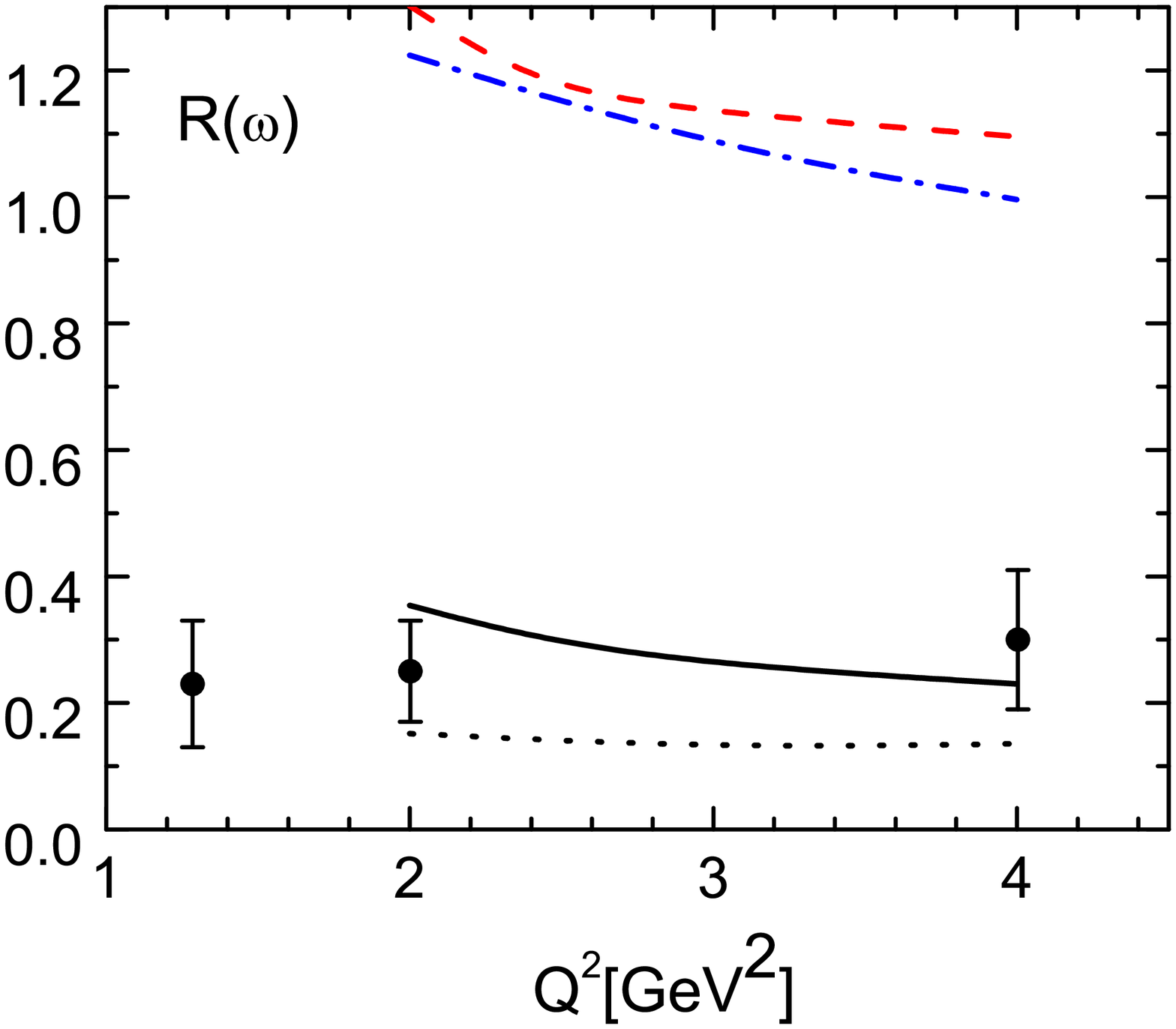}&
\includegraphics[width=7.2cm,height=5.8cm]{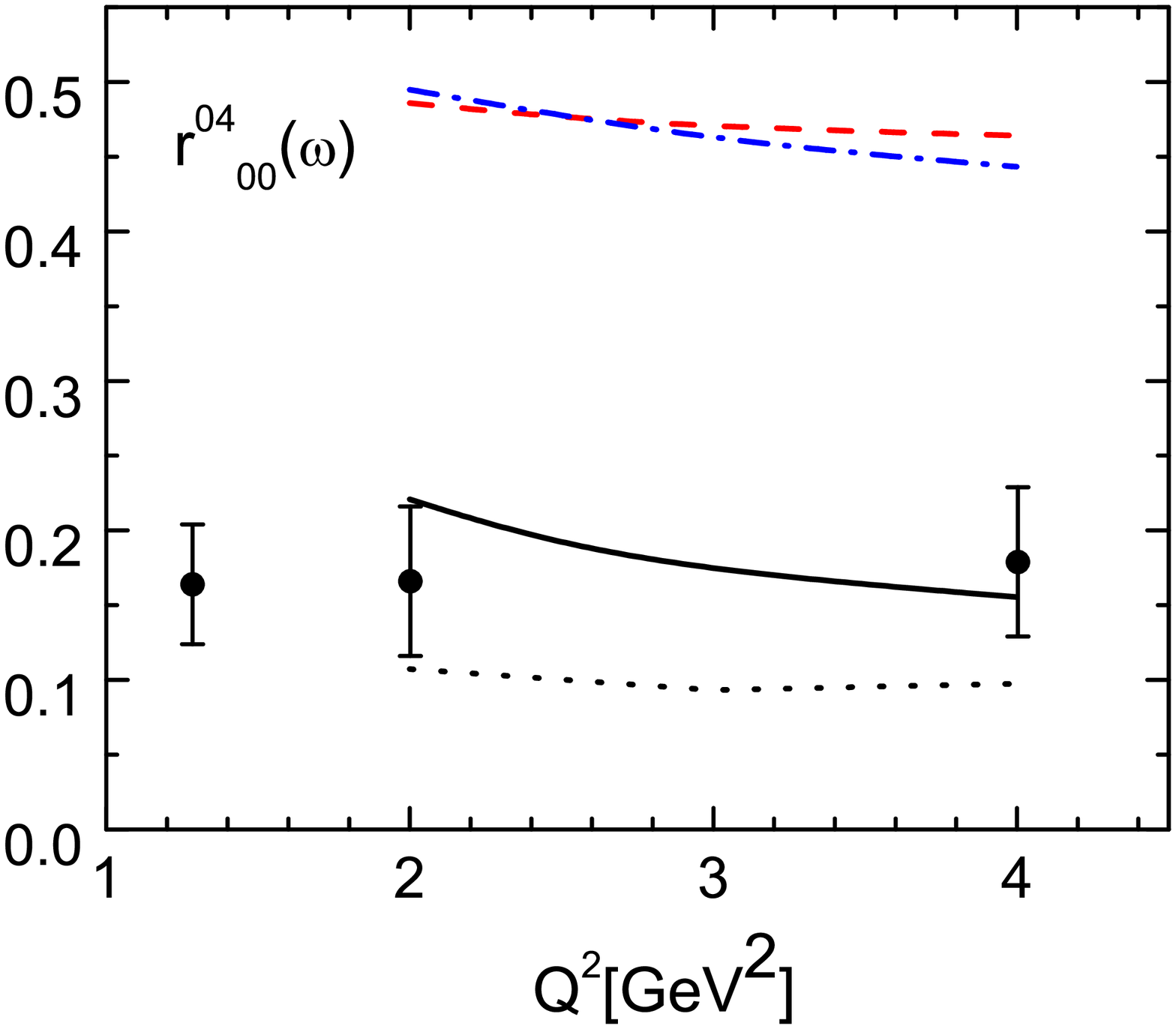}
\end{tabular}
\end{center}
\caption{Left: $R(\omega)$; Right: $r^{04}_{00}(\omega)$ at
HERMES: Black solid with PP, Red-dashed -without PP. Black
dotted-for $W=3.5 \mbox{GeV}$ (CLAS), Blue dashed-dotted for $W=8
\mbox{GeV}$ (COMPASS).}
\end{figure}

The SDME $r^{1}_{1-1}$ shows the difference of the natural and
unnatural parity contributions
\begin{equation}
r^{1}_{1-1} =- \mbox{Im}r^{2}_{1-1} =\frac{d \sigma^N(\gamma^*_T
\to V_T)-\sigma^U(\gamma^*_T \to V_T)}{2 d \sigma}.
\end{equation}
Results for this SDMEs for the $\omega$ production are shown in
Fig. 8, (left). We see that the PP effects are very strong here.
With PP we find $r^{1}_{1-1} \sim -0.2$ and without PP
$r^{1}_{1-1} \sim 0.2$. For the $\rho$ meson production results
are shown on Fig. 9, (right). The PP contribution is small here
and the results are close the to $\omega$ case without PP.

\begin{figure}[h!]
\begin{center}
\begin{tabular}{cc}
\includegraphics[width=7.2cm,height=5.8cm]{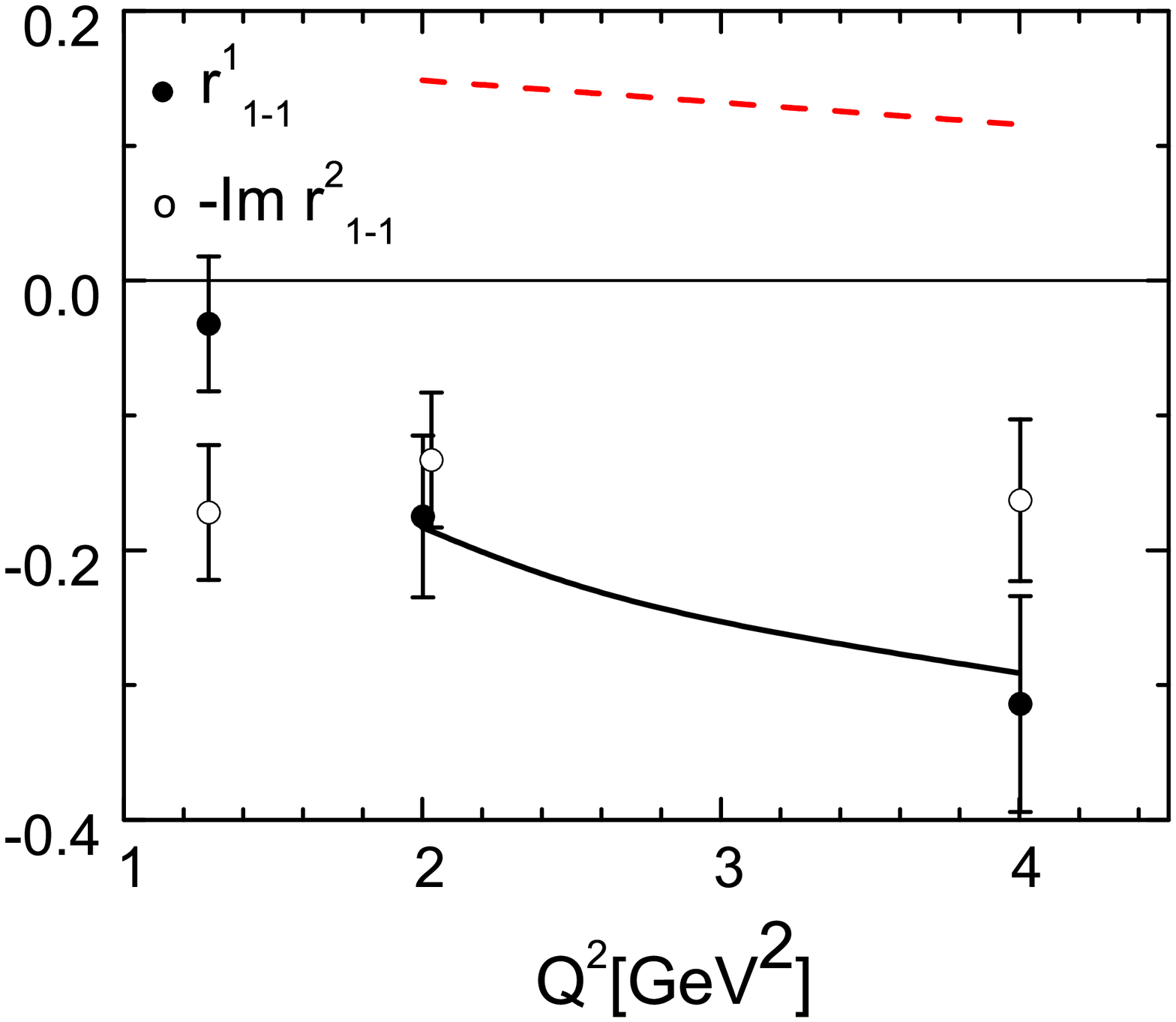}&
\includegraphics[width=7.2cm,height=5.8cm]{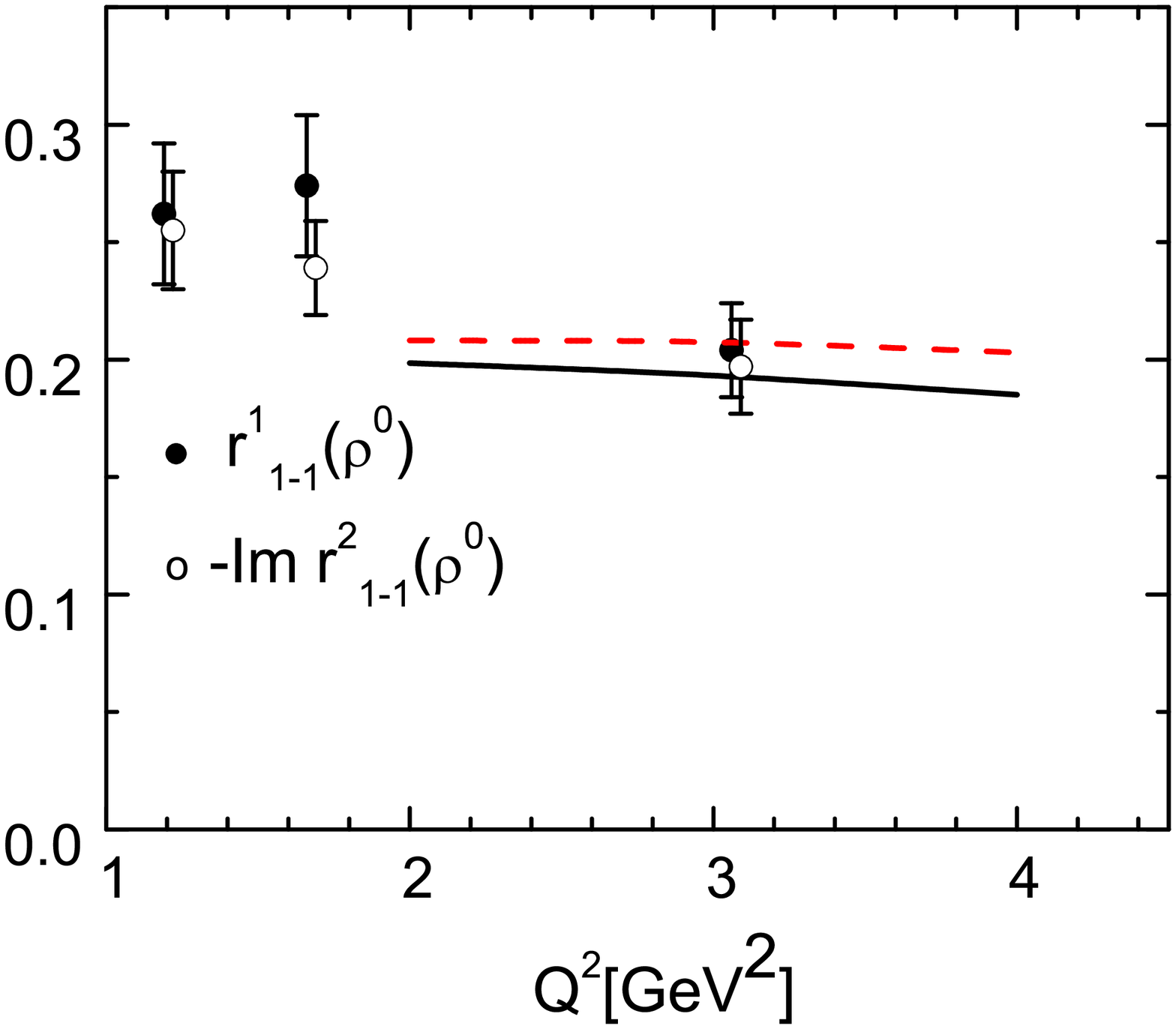}
\end{tabular}
\end{center}
\caption{Left: SDMEs for $\omega$ at HERMES: Right: SDMEs for
 $\rho$. At HERMES: Black solid with PP,
Red-dashed -without PP.}
\end{figure}
The SDME $\mbox{Im} r^6_{1-1}$ is proportional to interference of
two UP amplitudes
\begin{equation}
\mbox{Im} r^6_{1-1} \sim \mbox{Re} M^U(+-++) M^U(+-0+).
\end{equation}
Our model results show that the PP contribution to unnatural
parity amplitudes describes well the $Q^2$ and $t$ dependences of
$\mbox{Im} r^6_{1-1}$ SDME, see Fig. 9. Without PP this SDME is
equal to zero.
\begin{figure}[h!]
\begin{center}
\begin{tabular}{cc}
\includegraphics[width=7.2cm,height=5.8cm]{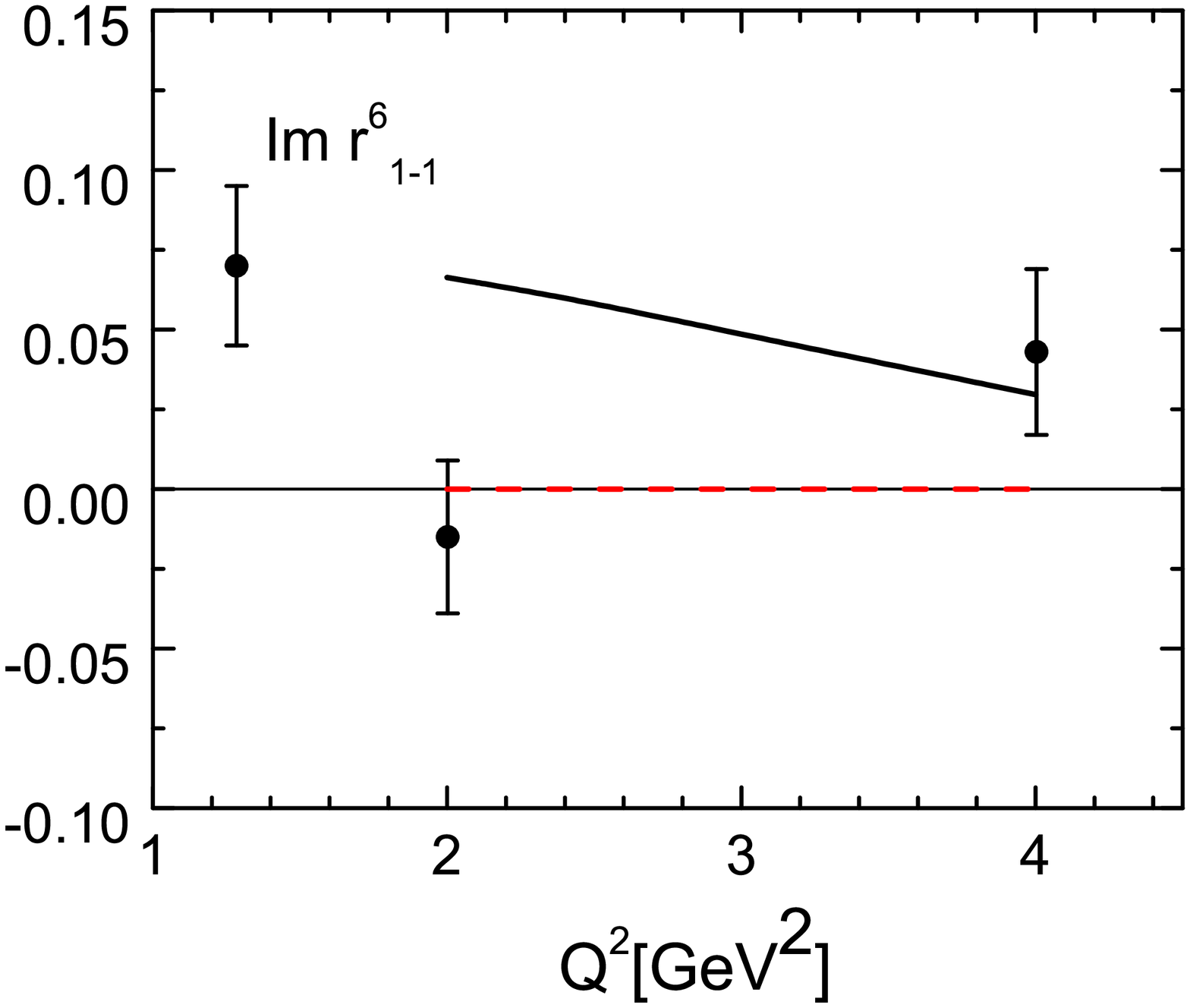}&
\includegraphics[width=7.2cm,height=5.8cm]{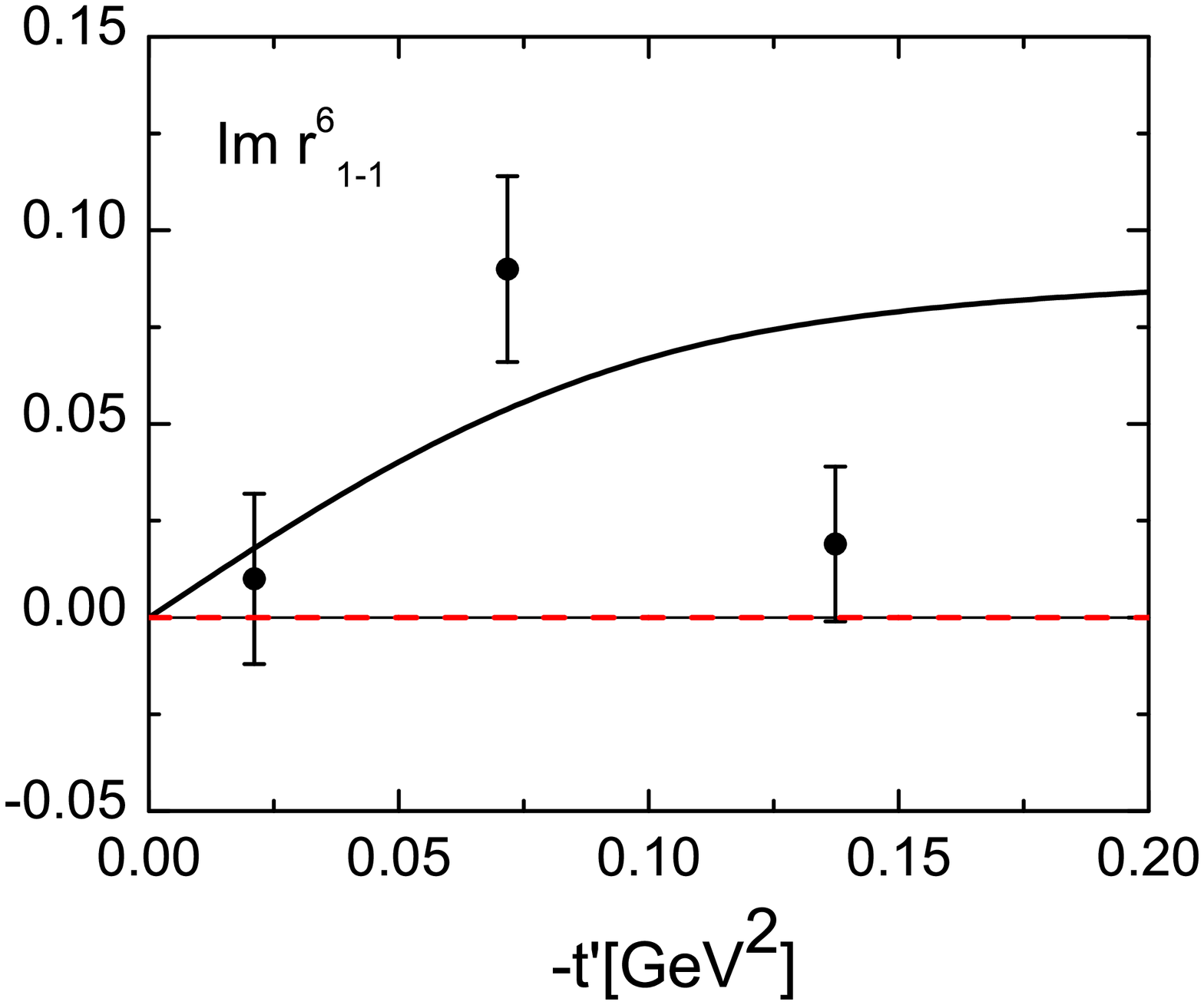}
\end{tabular}
\end{center}
\caption{SDME $\mbox{Im}$ $r^6_{1-1}$ for $ \omega$ at HERMES:
Left: $Q^2$ dependence of SDME ; Right:  $-t$ dependence of SDMEs:
Black solid with PP, Red-dashed -without PP}
\end{figure}
\section{Conclusion}
The exclusive electroproduction of vector and pseudoscalar mesons
was analyzed here within the  handbag approach where the amplitude
factorizes \cite{fact} into  the subprocess amplitudes and GPDs,
which contain information about the hadron structure.  The hard
subprocess amplitude is calculated within the $k_{\perp}$
factorization scheme \cite{sterman}. The results based on this
approach on the VM cross sections and various spin observables are
in good agreement with data at HERMES, COMPASS and HERA energies
at high $Q^2$ \cite{gk06}.

The role of transversity  $H_T$ and $\bar{E}_T$ GPDs in
leptoproduction of light mesons was investigated within the
framework of the handbag approach in \cite{gk09, gk11}. The
transversity GPDs in combination with twist-3 meson wave functions
occur in the amplitudes for transitions from a transversely
polarized virtual photon to a longitudinal polarized vector meson.
It was found that the transversity effects are essential in the PM
leptoproduction where they lead to large transverse cross
sections, which exceed substantially the leading twist
longitudinal cross section. There is an experimental indication
that the transversity effects in the PM production were likely
observed in CLAS \cite{bedl}. In the VM production, transversity
contributions were analysed \cite{gk13} in SDMEs and in
asymmetries measured with a transversely polarized target where
such effects are essential. The results are consistent with HERMES
and COMPASS data \cite{airap, autcomp13} on the $\rho^0$
production.

Using  GPDs from our analyses of hard meson leptoproduction we
investigated the $\omega$ SDMEs \cite{gk14} measured by the HERMES
Collaboration \cite{omega14}. It was found that PP give an
essential contributions to the $\omega$ production. The PP
contribution explains  the large unnatural-parity effects which
 exceed the natural parity contribution in the
$\omega$ production, as observed by HERMES.  For example the ratio
of the unnatural to the natural parity cross section  was found
for $\omega$ to be larger than unity instead of the  expected
small value. Results for the $\rho$ production were presented too.
The PP contribution in the $\rho^0$ production is much smaller
with respect to the $\omega$ case, which is consistent with
experiment \cite{airap}. Our results \cite{gk14} are in good
agreement with the HERMES experimental data \cite{omega14,airap}.

\bigskip

This work is supported  in part by the Russian Foundation for
Basic Research, Grant  12-02-00613  and by the Heisenberg-Landau
program.

\end{document}